# Celebrating the Physics in Geophysics

*Anthony B. Davis*

Los Alamos National Laboratory

LANL/ISR-2, Mail Stop B-244, Los Alamos, NM 87505

Voice: 1-505-665-6577 / Fax: 1-505-665-4414 / E-mail: adavis@lanl.gov

*Didier Sornette*

Institute of Geophysics and Planetary Physics
and Department of Earth and Space Science
University of California, Los Angeles, California 90095

Laboratoire de Physique de la Matière Condensée, CNRS UMR6622 and Université
des Sciences B.P. 70, Parc Valrose, 06108 Nice Cedex 2, France

**Abstract**

As 2005 —International Year of Physics— draws to an end, two physicists working primarily in geophysical research reflect on how geophysics is *not* an applied physics. Although geophysics has certainly benefited from progress in physics and sometimes emulates the reductionist program of mainstream physics, it has also enlightened the physics community concerning some of the generic behaviors of strongly nonlinear systems. Examples are the insights we have gained into the "emergent" phenomena of chaos, cascading instabilities, turbulence, self-organization, fractal structures, power-law distributions, anomalous scaling, threshold dynamics, creep, fracture, and so on. In all of the above, relatively simple models have captured the recurring features of apparently very complex signals and fields. The future of the intricate relation between physics and geophysics will be as exciting and productive as its past has been one of mutual fascination.

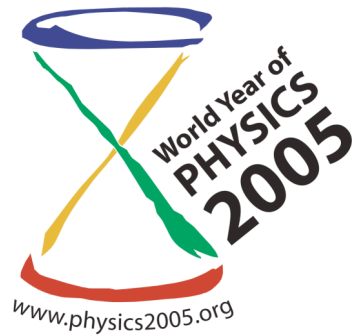

A. B. Davis and D. Sornette, **Celebrating the Physics in Geophysics**

UNESCO declared 2005 "International Year of Physics" in celebration of the centennial of Einstein's annus mirabilis when, junior clerk at the Swiss Patent Office in Berne, he published three papers that changed physics forever by 1) introducing Special Relativity and demonstrating the equivalence of mass and energy ($E = mc^2$), 2) explaining the photoelectric effect with Planck's then-still-new-and-controversial concept of light quanta ($E = h\nu$), and 3) investigating the macroscopic phenomenon of Brownian motion using Boltzmann's molecular dynamics ($E = kT$), still far from fully accepted at the time. Like surprisingly many others in AGU, we are trained as physicists but work primarily in geophysics. We enjoy thoroughly our research activity in atmospheric and solid-earth science respectively. In the following, we take a broad view to examine the past and future of the intricate and evolving relation between geoscience and physics.

Prefacing his *Principles of Philosophy*, the 17th-century French philosopher, mathematician and physicist Descartes described Philosophy as a tree rooted in Metaphysics, whose trunk is Physics with branches representing other scientific disciplines such as Mechanics or Medicine. This "Tree of Philosophy" is an extreme version of the general perception that Physics is the Queen of Sciences and that all others are essentially different flavors of Applied Physics. Phil Anderson, condensed-matter physicist and Nobel laureate, contended in his 1972 essay *More Is Different* [1], that particle physics and indeed reductionism have only limited ability to explain the world. He argued that reality is structured hierarchically, each level being independent, to some extent, of levels above and below: "At each stage, entirely new laws, concepts and generalizations are necessary, requiring inspiration and creativity to just as great a degree as in the previous one." Anderson noted "Psychology is not applied biology, nor is biology applied chemistry." We examine here how geophysics, far from being just another field of applied physics, has driven physics itself to innovate in deep and lasting ways.

Geophysical research is generally perceived as using basic scientific principles to explain observations while hypothesizing theories on the cause of still-unresolved structures and dynamics. In 1752, Franklin performed his famous kite-flying experiment, establishing that lightning is a naturally-occurring electric spark. In the 19th century, Tyndall's investigation of the radiative properties of gases contributed greatly to our understanding of how gases affect the heating and cooling of the atmosphere. Arrhenius, Nobel prize in Chemistry, presented groundbreaking work to the Stockholm Physical Society in 1895 "On the Influence of Carbonic Acid in the Air upon the Temperature of the Ground," a question still haunting Earth scientists and the broader community. Early in the 20th century, Serbian astrophysicist Milankovitch developed a mathematical theory of climate based on seasonal and latitudinal variations of solar radiation from varying Earth-Sun geometry (orbital eccentricity, obliquity, and precession). This theory still stimulates research in mathematics and physics, both applied and fundamental. Pioneering an interdisciplinary approach, Wegener wrote in 1915 one of the most influential and controversial books in the history of science, *The Origin of Continents and Oceans*, therein offering his theory of drifting continents and widening oceans to explain the evolution of Earth's geography. The theory waited almost 50 years to be rejuvenated as the theory of Plate Tectonics, established on the solid basis of modern geophysical data (rock magnetism, mantle convection and seismology). Recently, John Martin studied the basic chemical processes governing ocean life and proposed his "iron hypothesis:" sprinkling a small amount of iron into critical oceanic regions could generate extensive algae blooms that would remove so much carbon from the atmosphere that greenhouse heating could be reversed. These are just a few examples of how geophysicists have, like physicists, reduced complex phenomena to elementary processes —and even capitalized on their successes to propose bold geo-engineering projects— thus reinforcing the notion that geophysics is an applied field dependent on more general science.

However, there are also many instances where geophysics has directed scientists towards discoveries considered "fundamental," even by criteria satisfying the purest of physicists. In the early 20th century, Vilhelm Bjerknes, a founder of modern meteorology and weather forecasting, discovered circulation theorems that led him to a synthesis of hydrodynamics and thermodynamics applicable to large-scale atmospheric and oceanic motions. This opened the road to computation of the future state of the atmosphere by integrating primitive equations forward in time, starting from the observed initial state of the atmosphere. Bjerknes, his son Jacob and others established the mechanisms controlling the behavior of mid-latitude cyclones, whence the theory of air-masses and fronts. Motivated in part by the high pressures and shear stresses sustained by rocks inside the Earth, Bridgman in 1935 and later Enikolopyan did groundbreaking work on the effect of combined hydrostatic pressure and shear applied to a wide variety of materials. This research continues today at the interface between the quests for new properties of minerals under high pressures (up to megabars) and high temperatures (up to thousands of degrees) and for the physico-chemical properties of composite geological materials. These studies will potentially impact the location and quantification of oil or geothermal energy sources and structural investi-gations of silicate melts and glasses at the molecular level. Solid friction, formalized by Coulomb in 1773, has received a recent boost with the discovery of state- and velocity-dependent (Ruina-Dieterich) friction coefficients, motivated by its application to earthquakes. A vigorous research effort now attempts to unravel its microscopic physical origin in a variety of materials and to decipher its broad consequences on the dynamics of sliding and rupture in materials from atomic scales to faults hundreds of kilometers long. Wavelet analysis is the prevailing method to overcome the signal-cutting problem in Fourier analysis by using fully-scalable modulated windows. In its present form, the wavelet transform was proposed by Morlet in 1975 under the name "cycle-octave transform" while working for an oil company, with the goal of enhancing the resolution of seismic signals. This novel mathematical tool proved useful in data compression and denoising (leading to significant savings in storage and transmission costs), integration of partial differential equations, communication, image analysis, spatial statistics, and so on. The concept of fractals, developed by physicist/mathematician Benoît Mandelbrot in the 1960s, owes much of its inspiration to geoscience which, in turn, offers the richest and most diverse applications (topography, fault networks and earthquake ruptures, rocks, lightning, river networks, coastlines, patterns of climate change, clouds, etc.). The discovery in 1963 of deterministic chaos by Lorenz, an MIT meteorologist, essentially completed Poincaré's theory of dynamical systems. This breakthrough triggered over two decades of intense exploration in mathematics and physics to understand the mechanisms of chaos. Fractals and chaos are sweeping concepts widely used in non-equilibrium physics at large. These are just a few examples of how geoscience as influenced mainstream physics and enabled progress in its own endeavors (e.g., chaotic behavior in lasers, quantum chaos, multifractality in turbulence and statistical physics).

Chaos and fractals are now standard tools used in the broad field loosely called "complexity theory." They have played similar paradigmatic roles for the theory of complex systems as relativity theory and quantum mechanics did for physics in the first half of the 20th century, and DNA's double helix did for life sciences in its second half.

The still-evolving theory of complexity is the blueprint for a new way of doing science in which geosciences are both users and builders of fundamental principles. Indeed, the notion of complex systems and the importance of systemic approaches is pervasive in geoscience, arguably the best "laboratory" for exploring the following concepts: systems with a large number of interacting parts, often open to their environment, self-organizing their internal structures, and dynamics with novel and sometimes surprising ("emergent") macroscopic properties. Complex-system approaches see the whole as well as the parts and how they interact. Complexity manifests in linkages between space and time, often producing patterns on many scales and hierarchies of interactions, cascades —both direct (from large to small scales) and inverse (micro to macro)— and generation of fractal structures. Complex systems are typically sensitive to initial conditions. They may be robust with respect to certain types of perturbation, fragile with respect to others, and exhibit transitions from different classes of order to chaos, turbulence, or even complete randomness. Concrete examples of fundamental concepts in statistical physics inspired by the geosciences include block-spring models and self-organizing sandpile models for earthquakes and rupture under threshold dynamics [2].

Feeding the quest for new concepts and understanding, geoscience is witnessing an unprecedented accumulation of data with improved accuracy, frequency, and resolution. The main drivers are pressing societal problems associated with biospheric sustainability. For instance, NASA's EOS program is gathering new climatological data from space while DOE's ARM program has taken world-leadership in establishing ground-based climate observatories around the globe. Particularly challenging is the long-term monitoring needed to understand ramifications of temporal and spectral variability of solar irradiance, aerosols, humidity, clouds, and precipitation for global warming, especially in the sensitive cryosphere. The intricate processes of air pollution, ozone depletion, El Niño, and so on, are thus coming into sharper focus with NASA and ESA missions such as Terra, Aqua, Aura, and EnviSat. NSF, USGS and other agencies are launching EarthScope, a project for monitoring the lithosphere at spatio-temporal scales undreamed off until recently. EarthScope will stimulate the scientific community in quest for seismic and volcanic hazard control, earthquake risk management, mineral formation, and so on. The Earth's surface and interior will thus be probed by novel arrays of instruments and new discoveries are anticipated in the fundamental physics of earthquakes, volcanoes, mantle convection and tectonics. It is remarkable that Plate Tectonics, which unifies much of geoscience, was not fully accepted until the late 1960s, decades after physicists understood most of the atomic and sub-atomic world as well as the finer details of nuclear energy production in the Sun. This is not because geoscientists are any less clever than physicists, but lies in the fact that gathering and making sense of data up to planetary scales requires time, commitment, resources and ingenuity beyond those usually required for most well-controlled experiments carried out in a laboratory.

The new wave of intensive data acquisition is expected to deeply invigorate geoscience, and therefore physical science too.

Because it deals with such complex systems, geophysics is rather messy, with little or no experimental control. It shares this last attribute with astrophysics, and more specifically cosmology, since there is only one Universe and only one Earth. Unlike cosmology however, the first inkling of physics-based prediction will be immediately confronted with hard data. So geophysics is a world of paradox and struggle because models of complex natural systems cannot be validated in the usual sense. Indeed, model validation by comparison of some output with real-world data does not confirm the underlying scientific conceptualization. There are cases of models that made accurate, quantitative predictions, but later proved to be conceptually flawed [3]. Far from being an applied physics, geophysics has mutated the role of scientists who, in addition to remaining abreast in their traditional discipline, are now required to develop novel and still mostly undiscovered ways of validating their models and theories.

The challenge is to understand humanity's global environment, formed of interacting systems (atmosphere, ocean, land surface, biosphere) whose combined complexity exceeds that of any system previously considered by the physical, life or social sciences. We applaud AGU and EGU for paving the way with their respective Focus-Group and Section in "Nonlinear Geophysics" cutting across traditional geophysical disciplines, and for nurturing their connections with societies representing mainstream physics. Departments of Geography, Geophysics, Meteorology, Environmental, Atmospheric and/or Oceanic Sciences are already networking in support of Earth System Science —to become the ultimate interdisciplinary program when linked with social sciences. Maybe time has come to reach out to other Departments: Physics, Computer Science, Economics, Sociology, Anthropology … ?